\documentclass[prl,twocolumn,aps,showpacs]{revtex4}

\usepackage{graphicx}
\usepackage{amsfonts}
\usepackage{bm}

\def\Journal#1#2#3#4{{#1} {\bf #2}, #3 (#4)}

\begin{document}
\title{Electron Transport in Double Quantum Dot 
governed by Nuclear Magnetic Field 
 }

\author{Oleg N. Jouravlev and Yuli V. Nazarov}
\affiliation{
Kavli Institute of NanoScience, Delft University of Technology,
             Lorentzweg 1, 2628 CJ Delft,The Netherlands}

\begin{abstract}
We investigate theoretically electron
transfer in a doble dot in a situation where it
is governed by nuclear magnetic field: This has been
recently achieved in experiment \cite{QT}. We show how
to partially compensate the nuclear magnetic field
to restore Spin Blockade.
\end{abstract}
\maketitle

Much modern research is devoted to practical realization
of quantum manipulation
and computation (QMC).
Although QMC can be performed with convenient magnetic resonance
techniques \cite{NMR},  this necessary
involves macroscopically many identical spins.
The challenge is to do QMC with individual spin states,
e.g. those of localized electrons.
Remarkable experimental progress has been recently achieved
in preparation, manipulation and measurement of individual
spin quantum states in quantum dots. 
An important issue to resolve in this case 
is the spin measurement: 
to do this, one has to convert spin into charge and/or electric current \cite{Spintronics1,Spintronics2}.
Such conversion has been realized in
a single quantum dot \cite{fujisawa,hanson}. 
Other experiments were focused 
on the transport through two coupled quantum dots \cite{tarucha,marcus}.
Although such double dot is a more complicated system with
many additional processes influencing spin and charge transfer,
the advantage is an immediate access to spin-charge conversion.
In a double dot,
two electron spins can be entangled in the course of quantum manipulation
forming 
either symmetric spin singlet or antisymmetric 
triplet states.
This strognly affects electron transport giving rise to 
Spin Blockade of electron tunneling \cite{SpinBlockadeOld}. 
The quantum dots are commonly fabricated  in 
GaAs-based semiconductor heterostructures. 
The specifics of GaAs is a strong hyperfine interaction between 
electron and nuclear spins \cite{paget:82}. 
Therefore, the spin of an electron localized in a
a quantum dot can be strongly affected by the effective 
spin magnetic field $\bm B^{N}$ arising from random configuration
of many nuclear spins situated 
in the dot. This field helps the transitions between 
the components of spin doublet \cite{Siggi:02} as well as between
singlet and triplet states \cite{Siggi:01}.
It has been experimentally proved that the nuclear field not
only lifts Spin Blockade in a double dot but gives 
rise to time-dependent spin-driven oscillations of the current \cite{Tarucha2}. 
 Similar effects have been also observed in Quantum
Hall constriction \cite{Hirayama}.
The origin of the oscillations is the modulation of the current 
by nuclear field and feed-back of electron spin on nuclear spins
that results in their nutations \cite{Siggi}.
Since the nuclear field is random and hardly
controllable, its influence on the electron spin 
 significantly
complicates QMC. This  has motivated 
intensive research aimed to measure and to predict the effect of 
the nuclei on spins in quantum dots \cite{tarucha,QT,marcus}.

The study presented here has been stimulated
by very recent experiment in this direction \cite{QT}.
The advantage achieved in this experiment 
is the better control of
electron levels in the dots and, most importantly,
the possibility to control and tune
the tunnel coupling between dots in a wide range. Most interesting
results were obtained near the boundary of Coulomb diamond where
the states that differ by number of electrons in the dot 
are aligned in energy. The authors of \cite{QT} were able 
to demonstrate the order-of-value change of the current
by anomalously small external magnetic field $\bm{B}_{\rm ext} \simeq 5 mT$
that matches the nuclear field. By tuning the tunnel coupling,
they observe this effect in several different regimes.

We provide an adequate theoretical framework
for this experimental situation. We derive and solve density
matrix equation valid in the regimes of interest. We concentrate
on the fact that nuclear field randomly changes at time scale
bigger than that of electron dynamics but smaller than the measurement
time and therefore provides "frozen disorder" for electron spins. 
We achieve agreement with experiment in rather fine details.
Very imortant result of our analysis is that for any given configuration
of nuclear magnetic field there is always a value of
external magnetic field at which there is no current
--- stopping point.
This encourages us to speculate that the fast current measurement
in this setup can be used to measure and, via external feedback,
partly compensate the nuclear magnetic field. The setup would be
stabilized in the stopping point
where Spin Blockade is resored and  QMC is possibly enabled.

The charge configuration of the double dot is given by the number
of electrons in the left and right dot ($N_L$,$N_R$). 
The gate and bias voltages are tuned in the experiment to provide the following
transport cycle: $(1,1) \rightarrow (0,2) \mathop{\rightarrow}^{\Gamma_R} 
(0,1) \mathop{\rightarrow}^{\Gamma_L} (1,1)$.  
Two last processes are incoherent tunnel transitions
with electron transfer from the left and to the right lead, the tunneling
rates being $\Gamma_L$ and $\Gamma_R$ respectively.
The first transition may be coherent and is due to tunnel 
coupling $t$ between electron states in different dots.
If spin is taken into account, there are four possible quantum
states in $(1,1)$ configuration: a singlet $S$(1,1)
and there components of the triplet $T_{i}$(1,1).
As to $(0,2)$ configuration, only a singlet state $S_{\rm g}$(0,2)
participates in the transport: The triplet states of $(0,2)$ 
are much higher in energy owing to strong exchange interaction 
between two electrons in the same dot.
The essence of Spin Blockade is the spin selection rule
for $(1,1) \rightarrow (0,2)$. Provided spin is conserved,
there is no matrix element connecting
any triplet state $T$(1,1) and $S_{\rm g}$(0,2).
Therefore the transition does not take place, the system 
gets stuck in one of the triplet states and the current is blocked.

\begin{figure}[t]
\centering
\includegraphics[width=0.6\columnwidth]{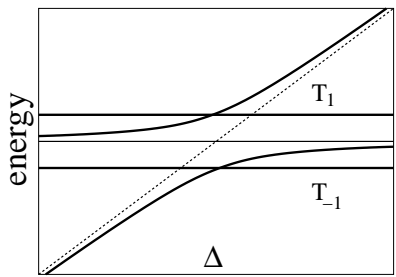}
\caption{Energies of the states with two electrons. 
The tunnel coupling between the dots mixes singlet states $S,S_g$ 
and does not influence triplet states (split by magnetic field).}\label{fig1}
\end{figure}

The part of the Hamiltonian for $(1,1)$ and $(0,2)$ configurations 
that conserves spin is presented 
in the triplet-singlet states basis ($T_i$,$S$ and $S_{\rm g}$) 
as
\begin{eqnarray}
\hat H_{0}=E \left( |S\rangle\langle S|+
\sum_i |T_{\rm i}\rangle \langle T_{\rm i}| \right )+
(E +\Delta) |S_{\rm g}\rangle \langle S_{\rm g}|+\nonumber \\
+ t( |S\rangle\langle S_{\rm g}|+ |S_{\rm g}\rangle\langle S|)\label{eq:Hamiltonian}
\end{eqnarray}
Here $\Delta$ is the detunning of $(1,1)$ and $(0,2)$ states 
linear in the gate and bias voltages. The experiments
were concentated at the edge of the Coulomb blockade diamond
where $|\Delta| \ll eV, E_C$. The tunnel coupling between the dots mixes
two singlets at $\Delta \simeq t$ but does not alter triplet states.
(Fig. \ref{fig1}) 

The leakage current in spin blockade regime 
can only arise from the spin-dependent interactions
that mix  singlet and triplet states. Theoretically, such interactions
can be caused by many mechanisms \cite{KhaetskiiNazarov}.
Experimenally, the most relevant one appears to be 
hyperfine interaction with nuclear spins. 
Since there are many nuclear spins interacting with 
an electron state  in each dot, their net effect can be presented 
in terms of  classical variables: effective fields
$\bm{B}^N_{L,R}$.(we measure fields in energy units) 
In the absence of net nuclear polarization,
these fields are random depending on a concrete configuration of 
nuclear spins \cite{Siggi:02}. Owing to central limit theorem,
the distribution of
both fields is Gaussian with
$\ll\bm{B}^{2}\gg \equiv B^2_N=E_n^2/N_{\rm eff} $, $E_n\approx 0.135meV$ for GaAs 
being the energy splitting induced by fully polarized nuclei, 
$N_{\rm eff}$ being the effective number of nuclei in the dot,
$N_{\rm eff} \simeq 10^{5-6}$ for typical dots. It is important for our
approach that nuclear fields change at time scale of nuclear spin relaxation
($ \simeq 1\ s$),
that is much bigger than any time scale associated with electron transport.
This is why they can be regarded as {\it stationary random} fields.
The electron spins inside the dots feel therefore 
effective stationary fields described by
\begin{equation}
\hat H_{\rm spin}= \bm{B}^N_{L} \cdot \bm{S}_L+\bm{B}^N_{R} \cdot \bm{S}_R
+B_{\rm ext} (S^z_L + S^z_R)
\label{eq:spinHamiltonian}
\end{equation}
$\bm{S}_{L,R}$ being the operators of the electron
spin in each dot and the external magnetic field is $\parallel z$. 
We rewrite this in triplet-singlet representation as 
\begin{eqnarray}
\hat H_{\rm spin}=(B_{\rm ext}+B_s^{z})\sum_i s_z^i|T_{\rm i} \rangle \langle T_{\rm i}|
+B_a^z|S \rangle \langle T_{\rm 0}|+\nonumber \\
\sum_{\pm}\left(\frac{B_{s}^x \pm iB_s^y}{\sqrt2}|T_{\rm 0} \rangle \langle T_{\rm \pm 1} |
+\frac{\mp B_{a}^x - iB_a^y}{\sqrt2}|S \rangle \langle T_{\rm \pm 1}|
+h.c.\right) \label{Hspin}
\end{eqnarray}
where $\bm{B}_{a,s}= (\bm{B}^N_L \mp \bm{B}^N_R)/2$ and $s_z^i=-1,0,1$ is the projection of the
spin of $|T_i\rangle$ state on $z$-axis.
We see that the sum of effective fields mixes and splits 
triplet components only. The difference 
of the fields mixes the spin singlet
$S$(1,1) and triplet $T$(1,1) states, this being
the source of leakage current.

The energy levels of the resulting Hamiltonian
$\hat H_{st}=\hat H_0 +\hat H_{\rm spin}$ 
are determined now not only by the tunneling $t$ and misalignment of the 
levels $\Delta$
but also by the fields, the corresponding energy scales can be comparable.
The mixing of the singlet and triplet in the eigenstates
of the Hamiltonian can be significant as well.
Already from analisys of this simple Hamiltonian
we can conclude that the current is absent if either
$\bm{B}_s \parallel \bm{B}_a$ or $\bm{B}_s \perp
\bm{B}_a$, since here $\bm{B}_s$ consists of the external and sum of nuclear magnetic fields.
To see this explicitly from (\ref{Hspin}),
let us choose $z$-axis in the direction of $\bm{B}_s$.
If $\bm{B}_s \parallel \bm{B}_a$, two triplet states
$|T_{\pm 1}\rangle$ are not mixed with the singlet.
If $\bm{B}_s \perp \bm{B}_a$, it is one state $|T_{0}\rangle$
that is not mixed. In both cases the system stucks in
one of the non-mixed triplet states resulting in no
current. Importantly, the stopping point
$(\bm{B}_s,\bm{B}_a)=0$ can be achieved at any
configuration of nuclear fields by adjusting
$B_{\rm{ext}}$.

To evaluate the current in general situation,  
we proceed with formulation of a suitable
density matrix approach first elaborated for double quantum dot 
in \cite{olddensity}. 
Current for the trasport cycle given is proportional to the probability 
to find a system in the state $S_{\rm g}$,
$I=e \Gamma_{\rm R}\rho_{\rm {S_g S_g}}$.
Although the transport involves 7 states, the probabilities
of $(1,0)$ doublets are readily expressed via other
probabilities. So the density matrix to work with
is spanned by five singlet-triplet states discussed.
Using the equations
of motion, we derive the equations for the
stationary density matrix ($d\hat \rho/dt =0$).
Five diagonal equations read
\begin{eqnarray}
\frac{1}{4}\Gamma_R \rho_{\rm S_g S_g}-
i\langle T_{i}| [\hat H_{\rm st},\hat \rho]|T_{i}\rangle=0\nonumber \\
\frac{1}{4}\Gamma_R \rho_{\rm S_g S_g}-\Gamma_{\rm in}\rho_{\rm SS}-
i\langle S| [\hat H_{\rm st},\hat \rho]|S\rangle=0\nonumber \\
-\Gamma_R \rho_{\rm S_g S_g}+\Gamma_{\rm in}\rho_{\rm S S}-
i\langle S_{\rm g}| [\hat H_{\rm st},\hat \rho]|S_{\rm g}\rangle=0\label{DensMatr}
\end{eqnarray}
where, motivated by experiment, we also
include inelastic transitions between $S$ and $S_g$ with the rate
$\Gamma_{in}$, $\Gamma_{in} \ne 0 $ if $\Delta < 0$.
The commutator terms include non-diagonal elements
of density matrix, so we also need 20 non-diagonal
equations,
\begin{equation}
-\frac{1}{2}(\Gamma^{j} +\Gamma^{i}) \rho_{\rm jk}-
i\langle j| [\hat H_{\rm st},\hat \rho]|k\rangle=0.
\label{DensMatr1}
\end{equation}
Here $j,k=T_i,S,S_{\rm g}$ number the five states basis,$j\ne k$.
The "rates" $\Gamma_{j}$ are zero for triplet states
and are $\Gamma_R$($\Gamma_{in}$) for $S_g$ ($S$).
To close the set of equations we use the normalization condition
for density matrix,
$\sum \rho_{jj}+\rho_{\rm S_g S_g}(1+\Gamma_R/\Gamma_L)=1,\,j=T_i,S$.


The solution gives the current for a given realization
of nuclear fields. Normally, one expects self-averaging
over different realizations at time scale of a single
measurement.
 Since nuclear relaxation times are large, this
point deserves some discussion.
In fact, raw data aquisition time in experiment
\cite{QT} was $0.1\ s$ per point, which is probably less
than  the relaxation time.
However, the raw data are noisy (see Fig. \ref{fig4})  due to both
instrumental noise and random changes of nuclear
fields. An accurate measurement requires, say, 50 data points,
this takes time much bigger than the relaxation time.
This leads us to two conclusions: (i) {\it smooth} part of experimental
data corresponds to the current {\it averaged} over
realizations,
(ii) a realistic (factor of 30) improvement of the measurement speed
and accuracy will allow to measure current for a given realization. 
So that, to compare our theory with experimental
results, we average the current obtained from the solution of equation set
(\ref{DensMatr}),(\ref{DensMatr1}) over Gaussian
distribution of fields.

\begin{figure}[t]
\centering
\includegraphics[width=0.85\columnwidth]{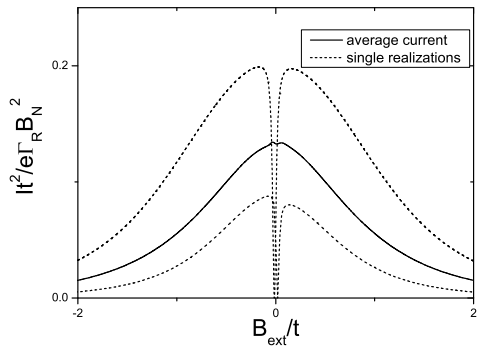}
\caption{Average current (solid) and that for
two random realizations of nuclear magnetic field in the limit of validity
of (\ref{current-small-BN})($t/B_N =50,\Delta/t=1$). Note stopping points at $B_{\rm{ext}} \simeq B_N$
seen for the realizations. }\label{fig2}
\end{figure}

Both solving and averaging can be easily done
numerically. To present the physics behind, we give
analytical results in two limiting cases.
The first, natural limit corresponds to small nuclear fieds,
$B_N \ll {\rm max}( t, B_{{\rm ext}})$.
In this case, the system is preferentially 
in one of the triplet states 
whose energies are $0, \pm B_s$.
It is convenient in this case to choose spin quantization axis 
along $\bm{B}_s$ and work with parallel and perpendicular components
of $\bm{B}_a$, $B_{a}^{\parallel,\perp}$ with respect to this axis.
The current reads
\begin{eqnarray}
\Gamma_R \ e/I = \left(\frac{t^2}{(B_{a}^{\parallel})^2} + 
 \frac{F(B_s)}{(B_{a}^{\perp})^2}\right);\; \nonumber \\ F(B_s) = t^2 +
 B^2_s(2+(B^2_s +\Delta^2)/t^2),
 \label{current-small-BN}
\end{eqnarray}
where the first term is due to transitions from $|T_0\rangle$
and the second due to transitions from $|T_{\pm 1}\rangle$.
As expected, the current stops if either $B_{a}^{\parallel}=0$
or $B_{a}^{\perp}=0$.
The average current in this limit 
\begin{equation}
I/e\Gamma_R = \left\{ \begin{array}{cc} 
B^2_N/15 t^2 & B \to 0 \\
B^2_N t^2/3 B^4 & B \to \infty
\end{array} \right.
\end{equation}
We plot the results in this limit for average curent
as well as for two arbirtaty realizatons of the field
(Fig. \ref{fig2}). The stopping points at $B_{\rm{ext}}\simeq B_N$ are
visible for realizations, while no features in average current are seen in this range

The alernative limit of big fields 
is achived provided $B_{{\rm ext}},B_N \gg \rm{max} (t, t^2/\Delta)$.
In this case, the system sticks in one of the four states (1,1)
with energies $\pm (B_L \pm B_R)$.
The current is determined by decay from these states and
reads 
\begin{equation}
I/e = \Gamma_R \frac{t^2 \left(\bm{n}_L \times \bm{n}_R\right)^2}{ 8  
(\Delta^2 + B^2_s )}
\label{current-big-BN}
\end{equation}
where $\bm{n}_{L,R}$ are unit vectors in the direction of $\bm{B}_{L,R}$.
At $B_{\rm ext} =0$ $I/e \simeq \Gamma_R t^2/B^2_N$ 
and it drops significantly at $B_{ext} \simeq \rm{max}(B_N, \Delta)$.

\begin{figure}[t]
\centering
\includegraphics[width=0.75\columnwidth]{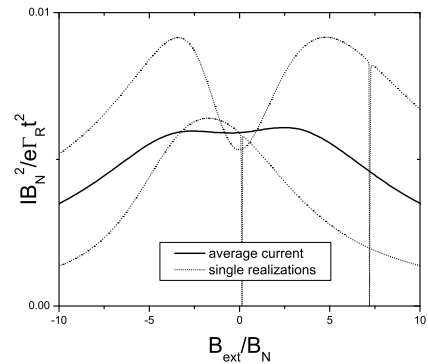}
\caption{Average current and that for
two relaizations in the limit of vadility of (\ref{current-big-BN})
($t/B_{N}=0.2, \Delta/t=50$).
Note narrow dips of the current at stopping points.}\label{fig3}
\end{figure}

\begin{figure}[t]
\centering
\includegraphics[width=0.85\columnwidth]{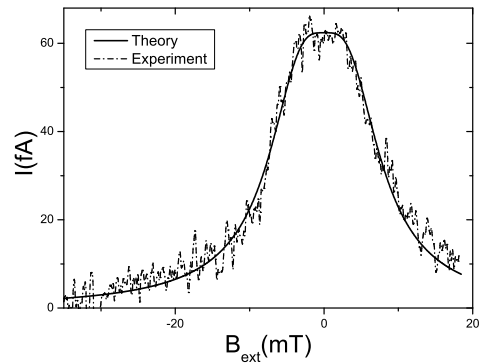}
\caption{Fit of experimental data \cite{QT} with "flat peak"
relation (\ref{current3}) gives $B_N=1.22\ mT$, $\Gamma_{in}=0.63\ MHz$ }\label{fig4}
\end{figure}

Eq. \ref{current-big-BN} seemingly contradicts to
our general statement giving non-zero current at $\bm{B}_s \perp \bm{B}_a$.
A fine point is that Eq. \ref{current-big-BN} is not valid in close 
vicinity stopping point where two of the four states
are degenerate. One has to take into account that this degerenacy
is lifted by coupling to $S_g$. As a result, the current develops a narrow
Lorenzian-shaped dip in the vicinity of the stopping point,
\begin{eqnarray*}
I/I_0 = \frac{(\delta B/B_w)^2}{1+ (\delta B/B_w)^2}; \\
\delta B = B_{ext} - B^{(0)}_{ext}, 
B_w \simeq \frac{t^2}{\Delta}
\end{eqnarray*}
Since the dip is narrow (see Fig. \ref{fig3}), it is washed away upon averaging.

The average current 
\begin{equation}
I/e\Gamma_R = \Biggl \{ \left.\begin{array}{cc} 
t^2 B_N^2/6B^2(\Delta^2+B^2) &\Delta,B \gg B_N \cr
t^2/12 \Delta^2 & \Delta \gg B_N \gg B_{ext} \cr
{\rm const}\; t^2/B^2_N & \Delta,B \ll B_N
\end{array}\right.
\end{equation}

We encounter a similar situation under conditions
where the four states are emptied by inelastic 
tunneling,
$\Gamma_{in}\gg \Gamma_R (t/\Delta)^2$ while their
splitting is determined by magnetic field, $B_N \gg \Delta_{ST}$.
It is experimentally confirmed that
this always takes place at sufficiently big negative $\Delta$.
The current is again an inverse of the sum of inverse partial rates
and reads
\begin{equation}
I/e = \Gamma_{in} \left(\bm{n}_L \times \bm{n}_R\right)^2
\end{equation}
The average current as a function of $B_{ext}$ becomes 
\begin{equation}
\langle I \rangle/e =\Gamma_{in} S(B_{{\rm ext}}/B_N); \label{current3}
\end{equation}
where
\begin{eqnarray}
S(x) \equiv 4/x^2 -6/x^4 
+\sqrt{2\pi} {\rm erfi}(x/\sqrt{2})(6/x^5\nonumber\\
-2/x^3) \exp(-x^2/2) - 
3\pi {\rm erfi}^2(x/\sqrt{2})\exp(-x^2)/x^6
\end{eqnarray}
In is interesting to note a special form
of this function: The graph of $S$ gives a peak with flat top, $S''(0) =0$.
This funbction provides an excellent fit to experimental data (Fig. \ref{fig4}),
those are impossible to fit with more conventional peak functions.
Such flat peaks are thus specific for the model in use and
provide strong support of its experimental validity.

In conclusion,
we have presented the theoretical framework for the electron
transport via a double quantum dot influenced and 
governed by nuclear magnetic field. Our approach is
based on density matrix equations and we achieve good agreement
with experiment \cite{QT} assuming averaging over realizations
of nuclear fields. An important feature which is yet to be observed
in the course of faster and more accurate measurement is the presence
of stopping points for any given realization of nuclear fields.

If one interprets the effect of nuclear magnetic fields in terms of 
spin coherence time, the results of \cite{QT} are discouraging if not
forbidding for QMC in GaAs quantum dot systems. The coherence time
estimated is just too short, $\simeq 10^{-7}\ s$. We speculate 
that the presence of stopping points can remedy the situation.
Faster current measurement would allow to characterize and,
with the aid of external feedback, partially compensate the nuclear fields
by stabilizing the system in the stopping point.

We are grateful to the authors of Ref.\cite{QT} for drawing our attention
to the topic, many usefull discussions and communicating
their results prior to publication. 
We acknowledge the financial support by FOM.

\end{document}